\documentclass[preprint,aps]{revtex4}

\usepackage{graphicx}
\usepackage{dcolumn}
\usepackage{bm}

\begin{document}

\title{Relevant energy scale in hybrid mesoscopic Josephson junctions}

\author{Franco Carillo$^{1,\dag}$, Detlef Born$^{1,2}$, Vittorio Pellegrini$^1$, Francesco Tafuri$^{1,2}$, Giorgio Biasiol$^3$, Lucia Sorba$^1$ and Fabio Beltram$^1$\\}
\address{
        $^1$NEST INFM-CNR and Scuola Normale Superiore, I-56126 Pisa, Italy \\
        $^2$Coherentia INFM-CNR Dip. Ing. dell'Informazione, Seconda Univ. di Napoli, 81031 Aversa,
        Italy\\
        $^3$Laboratorio Nazionale TASC-INFM-CNR,Area Science Park, I-34012 Trieste, Italy\\
        $^\dag$Corresponding Author, E-mail: franco.carillo@sns.it}
\date{22 July 2008}

\begin{abstract}
    Transport properties of high quality
Nb/semiconductor/Nb long Josephson junctions based on
metamorphic In$_{0.75}$Ga$_{0.25}$As epitaxial layers are
reported. Different junction geometries and fabrication
procedures are presented that allow a systematic comparison
with quasiclassical theory predictions. The impact of junction
transparency is highlighted and a procedure capable of yielding
a high junction quality factor is identified.
\end{abstract}

\pacs{Valid PACS appear here}
\maketitle

Superconductor/semiconductor hybrid devices are of much
interest not only for their potential for electronic-device
implementation but also as model systems for the investigation
of the phenomena regulating the conversion of supercurrent into
normal current at the interface through Andreev reflection
\cite{BTK}. By taking advantage of the increasing availability
of ultra pure nanoscale semiconductors, major advances on the
understanding of the microscopic nature of Josephson coupling
and the interplay between superconductivity and mesoscopic
effects can be expected \cite{IcQuant,IcKondoCNT,IcCB}. Recent
examples are the demonstration of the control of Josephson
current in diffusive InAs nanowires coupled to superconducting
leads (J-dot) \cite{IcCB} and the study of the interplay
between quantum interference effects and superconducting
correlation \cite{doh07,giazotto}.
Arguably, the main limiting factor for the exploitation of
hybrid systems in practical devices is the low value of the
junction quality factor $I_CR_N$ (i.e. critical current value
times normal resistance). In fact, with few exceptions
\cite{note,crestin}, this is the parameter normally referred to
in the literature \cite{Heida98,schapers98,kroemer}. It is
crucial to have the largest possible value of $I_CR_N$, since,
for instance, the maximum voltage amplification that can be
obtained from a Josephson-Fet \cite{Clark1980} or a J-Dot
\cite{IcCB} is proportional to $I_CR_N$. The values of $I_CR_N$
found in experimental works, however, are far from those
predicted by theory. This rises several issues about the actual
nature of these junctions and motivated transport analysis such
as the one reported here.

The reduction of $I_CR_N$ for semi/super Josephson junctions
was already discussed in terms of reduced dimensionality of the
normal conductor \cite{Kresin86}, low transparency of
semi/super interfaces \cite{Kuprianov88}, diffusive interface
\cite{Heida98}, and decoherence effects \cite{schapers98}. In a
recent work Hammer et al. studied
superconductor-normal-superconductor (SNS) junctions with
non-ideal S/N interfaces \cite{Hammer2007}. These authors
predicted that the Thouless energy ($E_{th}$) is replaced by the proximity
induced gap in the normal region as the relevant energy scale
governing, for instance, the temperature dependence of the
critical current (In a diffusive system with ideal transparent
interfaces $E_{th} = \hbar D/L^2$).

The objective of our work is two-fold: i) to investigate
experimentally the theoretical predictions on the impact of the
actual transparency of SN interfaces on SNS systems and ii) to
identify a processing strategy capable of yielding high quality
junctions. We shall examine various junctions characterized by
different transparencies and N regions of various length and
compare experimental Ic vs T curves for different fabrication
processes. From these curves, thanks to the knowledge of the
parameters of N material , we shall be able to analyze our data
in the frame of ref.~\onlinecite{Hammer2007} and identify the
process yielding the maximum interface transparency. In
particular, we report a semiconductor/superconductor interface
configuration for which $I_CR_N$ is close to $0.5$mV for a
junction length of $400$nm, among the best results ever
obtained for such hybrid devices.

\begin{figure}
\includegraphics[width=1.0\linewidth]{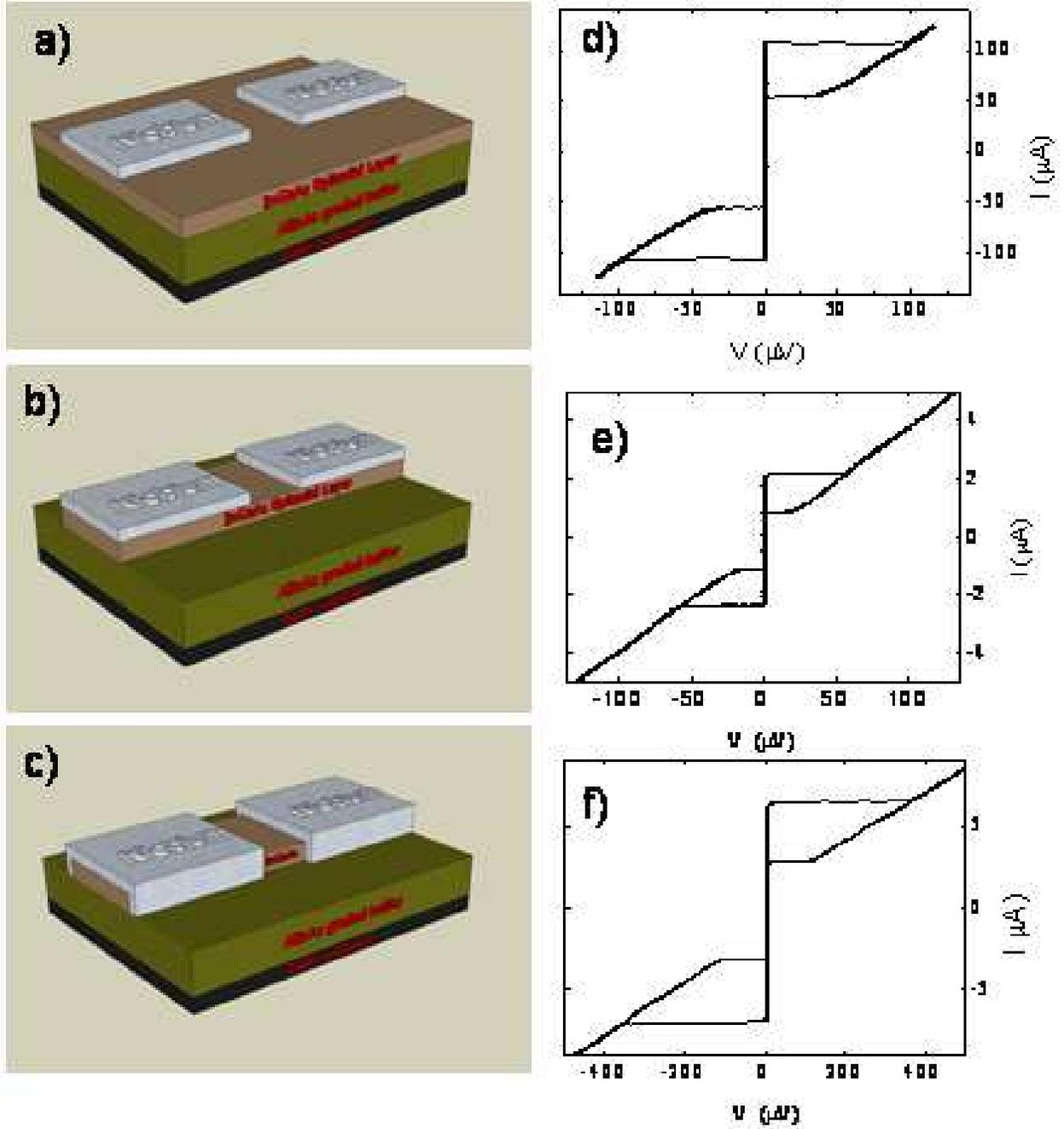}
\caption{\label{fig.schema} Panels a), b) and c) report schemes of type A, B, C junctions respectively:  GaAs substrate(black), AlInAs graded buffer(dark green), In$_{0.75}$Ga$_{0.25}$As epitaxial layer(brown), Niobium(gray). Panel d), e) and f) show I-V curves at $T=250$mK of junctions J1, J4, and J5 (see Table~\ref{table.eth}}
\end{figure}

In the hybrid devices of interest here, the normal conductor is
an epitaxial layer of In$_{0.75}$Ga$_{0.25}$As bulk-doped with
silicon. Thickness is in the 50nm-200nm range. Structures were
grown by molecular beam epitaxy (MBE) on a GaAs
$(100)$-oriented substrate \cite{tasc}. A sequence of
In$_{x}$Al$_{1-x}$As layers of increasing In content was first
deposited in order to ensure lattice matching with the
upstanding layer of In$_{0.75}$Ga$_{0.25}$As. Before Nb
deposition we performed a two-step surface cleaning of the
semiconductor consisting of a first wet removal of the native
oxide using diluted (1/50) HF solution and a subsequent vacuum
RF discharge cleaning at very low power in Argon. Nb was then
sputtered $in$ $situ$ during the same vacuum cycle and
electrodes were defined by lift-off. All Nb films were 85nm
thick. We shall call the device obtained with this process type
``A'' (Fig.~\ref{fig.schema}-a). Type ``A'' (Junctions J1 and
J2 reported in Table~\ref{table.eth}) were fabricated on a
200nm thick epilayer with charge density
$1.32\times10^{24}$m$^{-3}$ and mobility 0.69m$^2/Vs$ at $1.5$K
in the dark. A second type of device was fabricated on 50nm
thick epitaxial layers with charge density
$n=2.52\times10^{24}$m$^{-3}$ and mobility $\mu=$0.52m$^2/Vs$.
We defined the semiconductor geometry (junction width) by
employing a negative e-beam resist as a mask and subsequent wet
chemical etching in H$_2$SO$_4$/ H$_2$O$_2$/H$_2$O solution.
This type of device will be labeled``B'' and is showed in
Fig.~\ref{fig.schema}-b. A third type (``C'', shown in
Fig.~\ref{fig.schema}-c) of junction was obtained realizing the
semiconductor mesa before Nb deposition. A Ti mask is first
defined on the substrate by e-beam lithography, Ti thermal
evaporation and lift-off. In$_{0.75}$Ga$_{0.25}$As structures
are then defined by reactive ion etching. The last step is Nb
deposition using the same technique described for type ``A''.
For type ``C'' we employed the same epitaxial  layers of type
``B''. For all samples the transition temperature of Nb leads
is $T_c=8.7$K, from which, using $\Delta _{S} =
\Delta_{Nb}=1.9k_BT_c$, we calculate $\Delta_{Nb}=1.37$mV.

\begin{figure}
\includegraphics[width=1.0\linewidth]{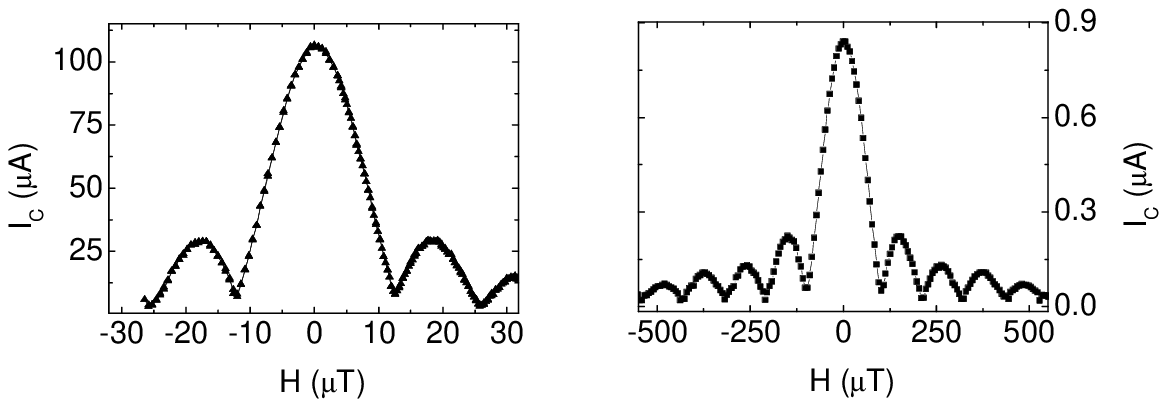}
\caption{\label{fig.fraun} Critical current (Ic) vs magnetic field (H) $T=250$mK for a type ``A'' junctions having $W=20\mu$m and $L=250$nm (left) and a type ``B'' junction with $W=5\mu$m and $L=900$nm (right)}
\end{figure}
Measurements were performed as a function of temperature down to
$250$mK in an $^3$He cryostat. Current and voltage leads were
filtered by RC filters at room temperature. A second filtering stage
consisted of RC + copper-powder filters thermally anchored at
1.5K-1.9K. The last copper-powder filter stage was at 250mK,
thermally anchored to the $^3$He pot. The shielding from magnetic
field was ensured by a combination of nested cryoperm, Pb and Nb
foils all of them placed in the measurement dewar and immersed in
liquid $^4$He.

In all junctions, with normal conductor lengths $L$ typically
ranging between 250nm and 1.1$\mu$m, we found a measurable
supercurrent, see Fig.\ref{fig.schema} panels d-f. For type A
structures, we measured the critical current in different
junctions made on the same chip (the same transparency was
assumed in this case). These junctions had different values of
electrode gap (1.1$\mu$m to 250nm) and consequently of barrier
length. Data are consistent with an exponential decrease of
Ic~vs.~L like in metallic SNS \cite{dubos}. The Fraunhofer-like
patterns shown in Fig.2 confirm the Josephson nature of the
zero voltage current. The period measured for junctions of
different widths and lengths is consistent with theoretical
expectations, which take into account flux focusing effects
\cite{ff}.
The critical current was evaluated using a $1\mu{V}$ criterion,
while $R_N$ was determined from a linear fit of I-V curves at
$V>3mV$.

For short ($\frac{\Delta_S}{E_{th}}\rightarrow0$) and tunnel
junctions the only relevant energy scale is the gap of the
superconductor $\Delta_S$ and at low temperatures $I_CR_N$
saturates at the value $\frac{1.326\pi\Delta_S}{2e}$. In the ideal case of
SN interfaces with zero resistance and very long SNS junctions
($\frac{\Delta_S}{E_{th}}>100$), $E_{th}$ determines both the
value at which $I_CR_N$ saturates at $T\rightarrow0$ and the
characteristic temperature of its exponential decrease. In this
limit $I_CR_N$ is not related to $\Delta_S$  but only to
$E_{th}$ through the expression $I_CR_N=bE_{th}$, where
$b=10.52$. While the ratio $\frac{\Delta_S}{E_{th}}$ decreases,
$b$ gets smaller. The complete $I_CR_N$ dependence on
$\frac{\Delta_S}{E_{th}}$ is reported in
Ref.~\onlinecite{dubos}. The proportionality between  $I_CR_N$
and $E_{th}$ for very long junctions was experimentally
demonstrated in the case of metallic SNS with transparent SN
barriers \cite{dubos}. For increasing SN interface resistance
Hammer et all. \cite{Hammer2007} showed that $I_C$ vs $T$
curves change their concavity from downward to upward and the
energy scale of their temperature decay is determined by an
effective Thouless energy ($E_{th}^{\star}$) which depends on
the ratio $r$ between the resistance of the SN interfaces and
the resistance of the normal conductor. In
ref.~\cite{Hammer2007} is shown that for
$\frac{\Delta_S}{E_{th}}\rightarrow \infty$ and large $r$
($r>10$) the following approximate relation holds:
$E_{th}^{\star}/E_{th}={Ar^B}/({C+r})$, where $A$, $B$, and $C$
are parameters which fit the numerical solutions and depend on
$\frac{\Delta_S}{E_{th}}$. We introduce $E_{th}^{\star}$ in our
discussion to have a single modeling parameter that has the
only purpose to account for barrier properties. As the
effective Thouless energy becomes smaller the value of $I_CR_N$
at low temperatures also becomes smaller and the decrease of
$I_C$ at high temperatures faster. A reduced effective value of
$E_{th}$ also implies a larger value of the ratio
$\frac{\Delta_S}{E_{th}}$, in other words the junction gets
longer as $r$ increases. In our paper, in agreement with
Ref.~\onlinecite{Hammer2007}, we link the temperature decay of
$I_C$ to $E_{th}^{\star}$ and show that the latter value also
determines the size of $I_CR_N$ at low temperature.
\begin{table*}
\caption{\label{table.eth} $(I_CR_N)_{th}$ is a theoretical
estimate made on the base of $E_{th}^{\star}$. Both $(I_CR_N)_{th}$ and  $(I_CR_N)$ are taken at $250$mK (see text).}
\begin{ruledtabular}
\begin{tabular}{cccccccccccc}
  & type  & W         & L     & J$_C$               &$E_{th}$         &$E_{th}^{\star}$          &$(I_CR_N)_{th}$     &$I_CR_N$        &$\frac{(I_CR_N)_{th}}{I_CR_N}$\\
  &       & [$\mu$m]  & [nm]  & [A/cm$^2$]          &[$\mu$eV]        &[$\mu$eV]                 &[mV]                &[mV]            &($T = 250$mK)\\
\hline
J1& A     & 20        & 250   & $2.68\times10^3$    & 545             &23                        &0.205               & 0.101          &2.0\\
J2& A     & 20        & 800   & $4.35\times10^2$    & 53              &10                        &0.074               & 0.030          &2.5\\
J3& B     & 10        & 800   & $1.18\times10^3$    & 52              &14                        &0.124               & 0.079          &1.6\\
J4& B     & 5         & 900   & $8.96\times10^2$    & 49              &16                        &0.149               & 0.059          &2.5\\
J5& C     & 0.45      & 400   & $2.84\times10^4$    & 284             &284\footnotemark[1]       &1.16                & 0.470          &2.5\\
J6& C     & 0.4       & 800   & $5.05\times10^4$    & 62              &62 \footnotemark[1]       &0.493               & 0.303          &1.6\\
\end{tabular}
\end{ruledtabular}
\footnotetext[1]{$E_{th}^{\star}$ has been taken equal to $E_{th}$, see text.}
\end{table*}
\begin{figure}
\includegraphics[width=1.0\linewidth]{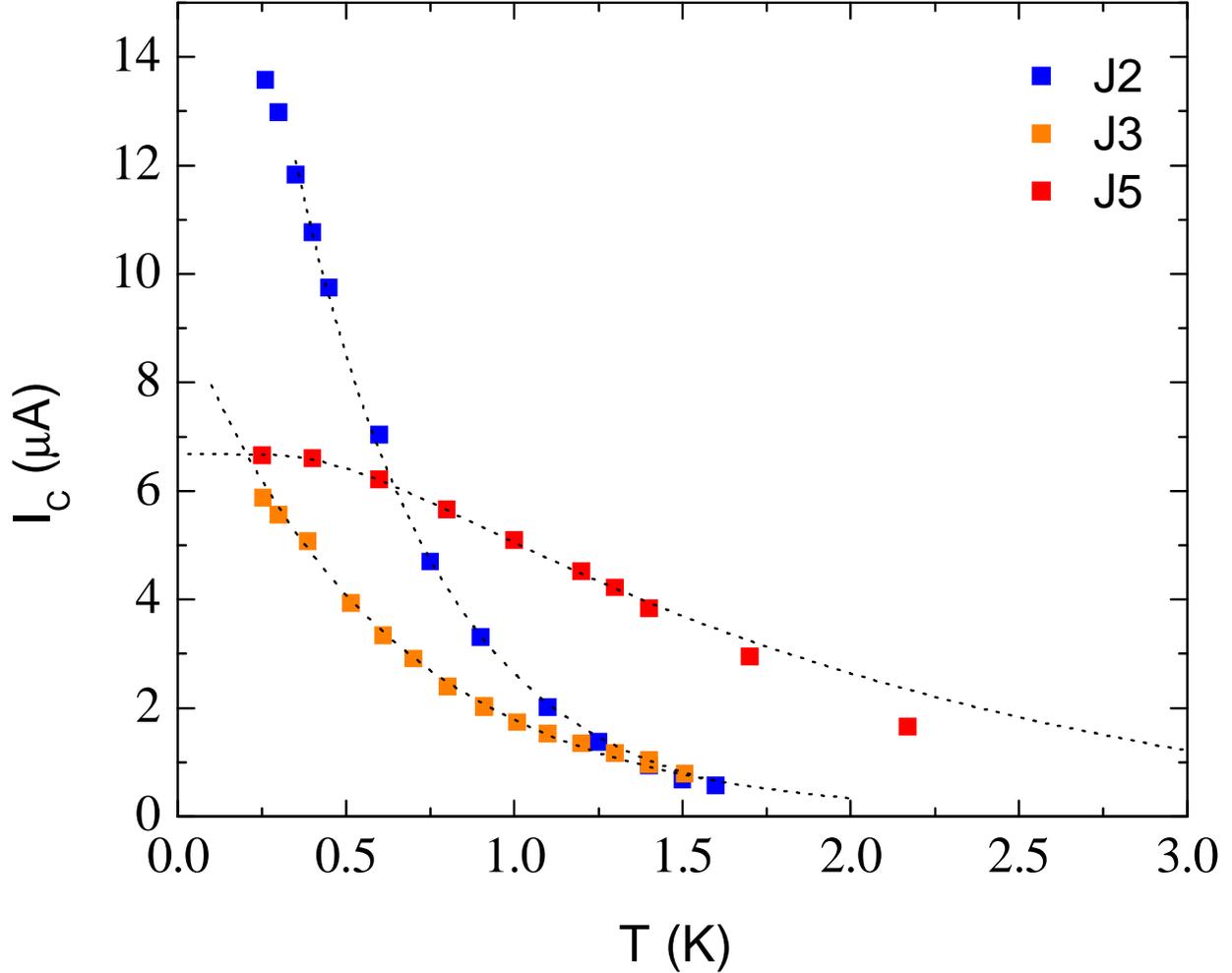}\\
\caption{\label{fig.ict} Ic vs T data points for junctions J2, J5, J3 (symbols) along with theoretical fit (dotted lines)}
\end{figure}

In order to provide an estimate of $E_{th}^{\star}$ we measured
$I_C$ vs. $T$ curves for all junctions. Three representative
results (filled points) are shown in Fig.3. $I_C(T)$ can be
well approximated by an exponential function for sufficiently
long junctions and high temperatures ($k_BT>5E_{th}$):
\begin{equation}\label{Icexp}
I_c(T)\propto\exp(-T/T^*),
\end{equation}
where $T^*$ is linked to the Thouless energy of the system by
$E_{th}^*=\frac{2{\pi}k_BT^{*}}{24}$ \cite{dubos}. We have used
Eq.~\ref{Icexp} to fit our data for all junctions of type A and
B as shown in Fig.3. The resulting values of $E_{th}^{\star}$
are reported in Table~\ref{table.eth}. For these devices
$E_{th}^{\star}$ is smaller than $E_{th}$, which can be
calculated from the relevant diffusion coefficients and
junction lengths.

The values of the ratio $\Delta_{Nb}/E_{th}^{\star}$ span from
60 to 170 and $k_BT{\gg}E_{th}^{\star}$ for all junctions in
the temperature range considered in the fit ($T{\geq}500$mK),
fully justifying the use of Eq.~\ref{Icexp} \cite{dubos}. Our
analysis of $I_C$ vs. $T$ shows that type A and B junctions
exhibit a behavior characteristic of much longer junctions with
respect to their actual geometric length.
$I_C$ vs. $T$ corresponding to junctions of type C is
remarkably different from those of other devices as shown in
Fig.3. The curve is concave downward and saturates at 250mK.
This curve shape is typical of junctions having SN interfaces
with low barrier resistance (see Figure 6 in
ref.~\onlinecite{Hammer2007}). For these devices $T<E_{th}/k_B$
in the whole temperature range considered (low temperature
regime) \cite{dubos,Hammer2007}, $I_C(T)$ can be approximated
by $I_C=(aE_{th}/eR_N)(1-ce^{-aE_{th}/3.2k_BT})$ \cite{dubos}
($R_N$ is $73.5\Omega$ for J5). An excellent fit is achieved
with an effective Thouless energy of $E_{th}^{\star} = 284$
$\mu eV$ and parameters $a=1.979$, $c=1.45$.
The quality of the fit, slightly deviating from the
experimental data only for $T>1.5K$, confirms that the value of
$E_{th}^{\star}$ is comparable to $E_{th}$ as opposed to cases
A and B. This suggest that the nature and/or the geometry of
super/semi interface characterizing type C can yield a higher
effective Thouless energy.

The reasons for which Type C interfaces have better performance
are not quite clear but are probably originated by a different
structure following the peculiar fabrication procedure. This
last includes Ti mask deposition, reactive ion etching, and
mask removal by HF solution. In this configuration an etched
wire of small width (450nm) is in contact with the
superconductor not only on the top of the smooth mesa surface,
like in cases A and B,  but also through the two sides, at the
rough surface of the etched side walls. It is however not
possible to ascribe with certainty the improvement of the
interface to the top contact, to the side contact or both. It
is widely accepted that the relatively poor quality of S/N
barriers is largely determined by reactive ion etching. Yet
this is funded on data for junctions whose SN side contacts
were taken on buried two dimensional electron gases
\cite{IcQuant,schapers98}. In those devices contacts were made
to the side of active layers less than $10$nm thick, which were
charge populated by modulation doping. Our junctions are based
on bulk doped surface layer and are therefore quite different
from most systems studied in literature. Bulk doped structures
like ours are more robust, as compared to modulation doped
quantum wells, to the formation of a charge depleted (dead)
layer at the two sides of the mesa after dry etching. We
believe that  the rough surface of the side walls (which is
still perfectly conducting in our case) promotes the adhesion
of metallic films thus improving effective contact area, and
transparency. An inhomogeneous contact can present areas with
different transparency: indeed the overall behavior of the
junction is determined by those areas having a large
transparency. Another mechanism worth future investigation is a
possible change in the chemistry of the top surface in the
epilayer due to deposition and subsequent removal of the Ti
mask. A chemical reaction at the surface between the mask and
the semiconductor could be favored, for instance, by local
heating of the mask during the RIE.

    In Tab.~\ref{table.eth} we calculate theoretical expected
$I_CR_N$ for J1-J6 after Ref. \onlinecite{dubos} on the basis
of $E_{th}^{\star}$. We find for all junctions, including J5
and J6, a factor that is around two between theoretical and
measured $I_CR_N$ \cite{notaRn}. This confirms the
effectiveness of the model described in
ref.~\onlinecite{Hammer2007}. Our data systematically show for
the first time that SNS systems with non ideal SN interfaces
can be characterized by an energy scale ($E_{th}^{\star}$)
capable to account for barrier resistance and transport
parameters simultaneously. The other interesting result is the
identification of a fabrication process for which
$E_{th}^{\star}$ is almost equal to $E_{th}$ (J5 and J6).
The agreement of $I_CR_N$ with existing theory is not perfect
for both high transparency junctions (J5 and J6), and low
transparency Junctions (J1-J4). A discrepancy with theory of
the same order was also found in some works based on metallic
SNS junctions \cite{Angers,courtois} and attributed to spurious
factors like the effective area of the contact \cite{dubos}.

Finally we note that J5 and J6 exhibit an $I_CR_N$ value very
close to the maximum obtainable for the given geometry and
material parameters for ideally transparent interfaces. To the
best of our knowledge similar results for
semiconductor/superconductor devices were obtained only for
junctions employing Si/Ge \cite{tinkham} and InAs nanowires
\cite{doh07}. We remark that, differently from the last two
cited works, our devices are based on a standard top-down
fabrication approach fully compliant with large scale
integration. We stress that J5 and J6 were fabricated on
different samples in different time. For J6 HF cleaning was
more concentrated (1/20 instead of 1/50) and the normal
conductor was ring shaped. To calculate the supercurrent
density for J6 we considered $W$ as the sum of of the width of
the two arms forming the ring. For the estimation of the
Thouless energy we considered $L$ equal to one half the average
circumference of the ring plus the gap between Nb electrodes
and the ring. High values of the characteristic voltage have
been obtained on similar structures produced with the same
fabrication process. These are mostly SNS junctions having ring
shaped normal conductor. Further details on SNS ring devices
will be reported elsewhere.

In conclusion we showed that the quality of semi/super
interface does play a crucial role in determining the value of
the effective Thouless energy in SNS junctions. The comparison
of different junction architectures shows that the reduction of
$I_CR_N$ is accompanied/mediated by the reduction of effective
Thouless energy. We determined a junction configuration
maximizing $I_CR_N$ and supercurrent density. This opens the
way to the realization of charge-controlled devices (J-dot)
with critical currents larger than $10$ nA.

We acknowledge useful discussions with A. Golubov, A. Tagliacozzo and P.
Lucignano. Financial support from EC within FP6 project
HYSWITCH (FP6-517567) is acknowledged.


\begin{thebibliography}{99}
\bibitem{note} The few works that make exception to this are mainly based on very short junctions
\bibitem{notaRn} In geometry ``A'' normal current can flow
    outside the close inter electrode area, where supercurrent
    is confined,  leading to a reduced value of $R_N$. Due to
    the favorable value of the ratio $W/L$ the understimate of
    $R_N$ is a small percent.
\end{thebibliography}
\end{document}